\documentclass[elsart,tightenlines,nofootinbib,superscriptaddress]{revtex4}

\usepackage{epsfig}
\usepackage{amsmath}
\input{epsf}
\usepackage{psfrag}
\usepackage{subcaption}
\usepackage[usenames,dvipsnames]{color}

\newcommand{\nn}{\nonumber \\}
\newcommand{\beq}{\begin{eqnarray}}
\newcommand{\eeq}{\end{eqnarray}}
\newcommand{\Slash}[1]{{\ooalign{\hfil/\hfil\crcr$#1$}}}

\begin{document}

\vspace*{2cm}
\title{Parton distribution function for the gluon condensate}

\author{Yoshitaka Hatta}
\affiliation{ Physics Department, Brookhaven National Laboratory, Upton NY 11973,  USA}

\author{Yong Zhao}
\affiliation{ Physics Department, Brookhaven National Laboratory, Upton NY 11973,  USA}

\begin{abstract}
\vspace*{0.5cm}
Motivated by the desire to understand the nucleon mass structure in terms of light-cone distributions,   
we introduce the twist-four parton distribution function $F(x)$ whose first moment is the gluon condensate in the nucleon. We present the equation of motion relations for $F(x)$ and discuss the possible existence of the delta function (`zero mode')  contribution at $x=0$.   We also perform one-loop calculations for  quark and gluon targets. 
\end{abstract}

\maketitle

\section{Introduction}

The hadronic matrix element of the  dimension-four scalar gluonic operator, or the `gluon condensate'
\beq
\langle P|F^{\mu\nu}F_{\mu\nu}|P\rangle, \label{condensate}
\eeq
is fundamentally important in hadron physics and beyond. This is primarily because the trace anomaly of QCD imparts mass to the nucleons and nuclei, hence to the visible universe, through the matrix element in Eq.~(\ref{condensate})~\cite{Jaffe:1989jz}. It thus plays a pivotal role in understanding the origin of the nucleon mass, a problem recently proclaimed by the National Academy of Science  \cite{nas} as one of the main scientific goals of the future Electron-Ion Collider (EIC) \cite{Aidala:2020mzt}.  
  However, the precise determination of  Eq.~(\ref{condensate}) turns out to be an  extremely challenging task.  A direct calculation from lattice QCD is notoriously difficult due to the vacuum quantum numbers of the operator involved (see a recent attempt \cite{Yang:2020crz}). Another possibility is that the matrix element can be probed experimentally in near-threshold quarkonium production  
\cite{Kharzeev:1998bz,Hatta:2018ina,Boussarie:2020vmu}. 

In this paper, we propose to study the partonic structure of the gluon condensate in Eq.~(\ref{condensate}) as a novel direction in the research of nucleon mass structure. Since this is a rather unusual proposal, to motivate the reader let us first draw an analogy to the study of nucleon spin structure. 
The Jaffe-Manohar sum rule  \cite{Jaffe:1989jz}
\beq
\frac{1}{2}=\frac{1}{2}\Delta \Sigma + \Delta G + L_q+L_g,  \label{jm}
\eeq  
tells how the total nucleon spin of $1/2$ is distributed among the helicity $\Delta \Sigma, \Delta G$ and orbital angular momentum $L_{q,g}$ of quarks and gluons. Each of these components can be expressed by the first moment of the corresponding parton distribution $\Delta\Sigma = \int dx \Delta q(x)$, $L_q=\int dx L_q(x)$, etc \cite{Hatta:2012cs}, where $x$ is the longitudinal momentum fraction. Such distributions are not only useful for extracting the moments from experiments, but also interesting in their own right, as they provide a more detailed, higher-dimensional description on the  spin structure.   
%In fact, the multi-dimensional partonic structure of hadrons has become an emerging paradigm especially in connection with the future Electron-Ion Collider (EIC). 

Returning to the problem of mass,  similarly to Eq.~(\ref{jm}), one can decompose the nucleon mass $M$ as 
\cite{Ji:1994av}
\beq
M=M_{kin}^q+M^{g}_{kin}+M_m + M_a, \label{massdec}
\eeq
where $M_{kin}^{q,g}$ are the kinetic energies carried by quarks and gluons, $M_m \sim \langle P|\bar{\psi}\psi|P\rangle$ is the contribution from the nucleon sigma term, and $M_a$ is from the gluon condensate Eq.~(\ref{condensate}). As in the case of spin decomposition, one naturally asks how partons with a given momentum fraction $x$  contribute to the four components in Eq.~(\ref{massdec}). For kinetic energy, this can be quantified by noticing that $M_{kin}^{q,g}$ are related to the second moment of the ordinary parton distribution functions (PDFs). One then sees that $M_{kin}^q$ is dominated by valence quarks at large-$x$. Gluons tend to have smaller $x$ values, but because there are so many of them, $M_{kin}^g$ can become sizable. On the other hand,   
%Despite the similarity to Eq.~(\ref{jm}), one seldom asks the question of how partons with a given value of $x$ contribute to the four components.  For the kinetic energy, perhaps the reason is because the answer : The kinetic energy is dominated by large-$x$ valence partons. This can be quantified by noticing that $M_{kin}^{q,g}$ are related to the second moment of the ordinary parton distribution functions (PDFs). 
 regarding the remaining two entries $M_{m,a}$, most of the work done so far has been limited to `zero-dimensional' physics. While the parton distribution function for $M_m$ does exist in the literature, called $e(x)$, its connection to hadron masses is not often emphasized. For $M_a$, the corresponding $x$-distribution  was almost nonexistent until very recently when related distributions were briefly mentioned in \cite{Ji:2020baz}. In principle, it is a  simple matter to write down the twist-{\it four} distribution 
%For $M_m$, one can define the twist-three distribution $e(x)$ 
%\beq
%e(x) \sim \int dz^- e^{ixP^+z^-}\langle P|\bar{\psi}(0)\psi(z^-)|P\rangle, \qquad \int dx e(x) \sim M_m.
%\eeq  
\beq
F(x) \sim\int dz^- e^{ixP^+z^-}\langle P|F_{\mu\nu}(0)F^{\mu\nu}(z^-)|P\rangle, \qquad \int dx F(x) \sim M_a.
\eeq
Together with the twist-two PDFs and $e(x)$, this provides a complete set of parton distributions for the nucleon mass structure.  
%To our knowledge, the property of $F(x)$ has not been investigated in the literature. 

In this paper, we present the first analysis of $F(x)$. We use the QCD equation of  motion to reveal its multi-partonic nature of the distribution. We then  present  one-loop calculations of $F(x)$ for quark and gluon targets. Particular attention is given to the question of whether $F(x)$ contains the delta function $\delta(x)$.  The (non)existence of $\delta(x)$ in $e(x)$ has been a subject of debate in the literature. We shall see that the discussion is entirely analogous for $F(x)$.  We shall present both  model-independent and model-dependent arguments in favor of the existence of the delta function.

\section{Chiral-odd twist-3 distribution $e(x)$}

Before introducing the twist-four gluon distribution $F(x)$, we first give a review of the twist-three, chiral-odd quark distribution $e(x)$. Our purpose is mostly to emphasize the similarity to $F(x)$ studied in the next section, but the present section also contains some original discussions.

 $e(x)$ is defined by 
\beq
e_q(x)= \frac{P^+}{2M}\int \frac{dz^-}{2\pi}  e^{ixP^+z^-}\langle P|\bar{\psi}_q(0) W[0,z]\psi_q(z^-)|P\rangle,
\label{e}
\eeq
where $M$ is the proton mass and $W$ is the straight Wilson line along the light-cone which makes the nonlocal operator gauge invariant. The distribution is defined for each quark flavor $q$ with mass $m$.  
The first and second moments are proportional to the nucleon sigma term and the number of valence quarks $N_q$, respectively
\beq
\int dx\,  e_q(x) = \frac{\langle P| \bar{\psi}_q\psi_q|P\rangle}{2M}, \qquad 
\int dx\, x e_q(x) = \frac{m_q}{M}N_q. \label{nq}
\eeq
 In what follows, we shall omit the subscript $q$ for simplicity.
By using the equation of motion and Lorentz invariant relation one can write  \cite{Ji:1993ey,Kodaira:1998jn,Efremov:2002qh,Pasquini:2018oyz}
\beq
e(x)= e_{sing}(x) + e_{tw3}(x) + e_{mass}(x),
\eeq
where $e_{sing}$ is proportional to the delta function at $x=0$,
\beq
e_{sing}(x) = \frac{\delta(x)}{2M}\langle P|\bar{\psi}\psi|P\rangle. \label{sing}
\eeq 
$e_{mass}(x)$ is related to the twist-two quark distribution $q(x)$ as 
\beq
e_{mass}(x) = \frac{m}{M} \left(\frac{q(x)}{x} - \delta(x) \int dx' \frac{q(x')}{x'}\right). \label{mass}
\eeq
Clearly, 
$\int dx\ e_{mass}(x)=0$.  
%\qquad \int dx xe_{mass}(x)=  \frac{m}{M}N_q, \nn
% \int dx x^2 e_{mass}(x)= \frac{m}{M} \int_0^1 dx x(q(x)+\bar{q}(x)) = \frac{m}{M}A_q
%\eeq
The `genuine twist-three' distribution $e_{tw3}(x)$ also has a delta function at $x=0$,
\beq
e_{tw3}(x)= \int dy \frac{\Phi(x,y)}{x(x-y)} - \delta(x) \int dx' dy' \frac{\Phi(x',y')}{x'(x'-y')},
\eeq
where
\beq
\Phi(x,y) =\frac{1}{2M} \int \frac{dz^-}{2\pi}\frac{dw^-}{2\pi} e^{ixP^+z^- +i(y-x)P^+w^-}
 \langle P|\bar{\psi}(0)W[0,w]\sigma^{+\mu}gF^{+}_{\ \mu}(w^-) W[w,z]\psi(z^-) |P\rangle,
\eeq
 is the quark-gluon-quark mixed distribution.  [Our sign convention for the QCD coupling is such that the covariant derivative reads $D^\mu = \partial^\mu + igA^\mu$.]  It is easy to see that   
\beq
\int dx\ e_{tw3}(x)=0, \qquad \int dx\ x e_{tw3}(x)=0 .
\eeq
The latter relation follows from the property $\Phi(x,x')=\Phi(x',x)$.
On the other hand, the third moment of $e_{tw3}(x)$ is nonvanishing 
\beq
\int dx\ x^2e_{tw3}(x) = \frac{1}{4M(P^+)^2} \langle P|\bar{\psi}\sigma^{+\mu}gF^{+}_{\ \mu}\psi|P\rangle.
\eeq
This matrix element is related to the electric dipole moment of the nucleon \cite{Seng:2018wwp}. 
One thus arrives at the relation
\beq
e(x) &=& \frac{m}{M} \frac{q(x)}{x} +  \int dy \frac{\Phi(x,y)}{x(x-y)} \nn 
&& + \delta(x) \left[  \frac{\langle P|\bar{\psi}\psi|P\rangle}{2M}  - \int \frac{dx'}{x'}\left( \frac{m}{M} q(x') +\int  dy' \frac{\Phi(x',y')}{x'-y'}\right) \right].  \label{ar}
\eeq

There have been  discussions about the nature of the delta function terms, or `zero modes', in Eq.~(\ref{ar}). 
Ref.~\cite{Efremov:2002qh} argues that the sum rule $\int dx e(x)\propto \langle P|\bar{\psi}\psi|P\rangle$ is of `no practical use' because the only contribution comes from the delta function at $x=0$ which experiments cannot access. [Remember that $\int dx\ e_{mass}(x)=\int dx\ e_{tw3}(x)=0$.] The presence of `zero modes' signifies the nonperturbative dynamics of QCD which leads to confinement and the generation of hadron masses. On the other hand, one can make an argument that the delta function may actually be absent. This is indeed the case in the naive parton model owing to the Weisberger relation \cite{Weisberger:1972hk} which in the modern notation reads \cite{Brodsky:2007fr}\footnote{Here is a quick derivation of the Weisberger relation in the parton model.
\beq
\langle P| \bar{\psi}\psi|P\rangle_{proton} &=& \int_0^1 \frac{dx}{x} (q(x)+\bar{q}(x))  \langle xP|\bar{\psi}\psi|xP\rangle_{quark}
\nn
&=& \int_0^1 \frac{dx}{x} (q(x)+\bar{q}(x))  \bar{u}(xP)u(xP) \nn
&=&2m  \int_0^1 \frac{dx}{x} (q(x)+\bar{q}(x))  .
\eeq 
The factor $1/x$ comes from the relativistic normalization of states. 
}
\beq
\frac{\partial M}{\partial m} = \frac{\langle P|\bar{\psi}\psi|P\rangle}{2M} = \frac{m}{M} \int_{-1}^1 \frac{dx}{x} q(x) =  \frac{m}{M} \int_0^1 \frac{dx}{x} (q(x) +\bar{q}(x)). \label{wei}
\eeq 
Since the genuine twist-three physics is absent in the parton model,  the expression inside the square brackets in Eq.~(\ref{ar}) vanishes. 
%The left hand side is nothing but the nucleon sigma term
%\beq
%\frac{\partial M}{\partial m} = \frac{\langle P|\bar{q}q|P\rangle}{2M}
%\eeq  
 However, Eq.~(\ref{wei}) is obviously problematic because the $x$-integral does not converge in real QCD. 
Going beyond the parton model, very recently the authors of \cite{Ma:2020kjz} claim to have shown 
 that the coefficient of the delta function vanishes exactly in full QCD. 
Their proof starts by writing $\bar{\psi}\psi=\bar{\psi}_+\psi_- + \bar{\psi}_-\psi_+$ where  $\psi_{\pm}= \frac{1}{2}\gamma^\mp \gamma^\pm \psi$ are the so-called `good' and `bad' components of the quark field, respectively. It is often stated in the literature that $\psi_-$ is not an independent field. Using the equation of motion one can write
\beq\label{eom2}
2iD_- \psi_- = (i\gamma^iD_i+m)\gamma^+\psi_+\,,
\eeq
where $i=1,2$.
The general solution to Eq.~(\ref{eom2}) is 
\begin{align}
\psi_-(z^-) &=\frac{1}{2i} \int dz'^- K(z^--z'^-)W[z^-,z'^-]  (i\gamma^iD_i+m)\gamma^+\psi_+(z'^-) + \int dz'^-W[z^-,z'^-]\psi_-^0(z'^-) \nn
&= \frac{1}{2}\int \frac{dx}{2\pi} K(x)\int dz'^- e^{-ixP^+(z^-  -z'^-)} W[z^-,z'^-]  (i\gamma^iD_i+m)\gamma^+\psi_+(z'^-) + \int dz'^-W[z^-,z'^-]\psi_-^0(z'^-) , \label{eom}
\end{align}
where $K(z^-)$ is the Green function subject to the boundary condition. Common choices are $K(z^-)=\theta(z^-)$, $-\theta(-z^-)$ and $\frac{1}{2}\varepsilon(z^-)=\frac{1}{2}(\theta(z^-)-\theta(-z^-))$. In momentum space, $K(x)=\frac{1}{x+i\epsilon}$, $\frac{1}{x-i\epsilon}$ and ${\rm P}\frac{1}{x}$, respectively. (P denotes the principal value.)  $\int \psi_-^0$ is not constrained by the equation of motion and should be treated as an independent field.  It  is essentially the zero mode  as it involves an unconstrained integration over $z'^-$ (up to a gauge rotation). In the literature, this term is routinely neglected when one works in the light-cone gauge $A^+=0$ and specifies the boundary condition at $z^-=\pm \infty$ in order to quantize the theory. Often the antisymmetric boundary condition, corresponding to P$\frac{1}{x}$, is employed  (see, e.g., \cite{Kogut:1969xa}), but this implicitly assumes the subtraction of the zero mode. While such a procedure may be justified for most purposes, like doing perturbation theory and computing the S-matrix, it may not capture the long-distance physics responsible for the generation of hadron mass.

Ref.~\cite{Ma:2020kjz} only kept the first term of Eq.~(\ref{eom}) with the advanced boundary condition $K(z^-)=-\theta(-z^-)$ and showed that the coefficient of the delta  function in Eq.~(\ref{ar}) vanishes exactly. Actually, it does not matter which boundary condition is adopted, because in the end only the combination $K(x)+K^*(x)=2{\rm P}\frac{1}{x}$ appears  in the sum $\langle \bar{\psi}_+\psi_- \rangle+\langle \bar{\psi}_-\psi_+\rangle=\langle \bar{\psi}_+\psi_- \rangle + (\langle \bar{\psi}_+\psi_- \rangle)^*$. 
However, the  $\psi_-^0$  
 term does not cancel  and leads to a nonvanishing coefficient 
\beq
 && \frac{\langle P|\bar{\psi}\psi|P\rangle}{2M}  - \int \frac{dx'}{x'}\left( \frac{m}{M} q(x') +\int  dy' \frac{\Phi(x',y')}{x'-y'}\right) \nn 
 && \qquad = \frac{1}{2M} \int dz^-\langle P| \bar{\psi}_+(0) W[0,z^-]\psi_-^0(z^-)+\bar{\psi}_-^0(z^-)W[z^-,0]\psi_+(0)|P\rangle.
\eeq

There is vast literature on the zero mode problem in light-front quantization (see, e.g., \cite{Nakanishi:1976vf} and  reviews \cite{Yamawaki:1998cy,Brodsky:1997de}). 
One might argue that in continuum theory the zero mode has no effect on physical observables because it has  measure zero in the path integral sense. On the other hand, entirely neglecting the zero mode causes serious inconsistencies  such as the lack of Lorentz invariance \cite{Nakanishi:1976vf}. This is still an open problem, and discussions of the quark and gluon condensates cannot be complete without a full consideration of the zero mode. For the moment, it seems to us that    
the coefficient of the delta function is likely nonvanishing, and can be determined only nonperturbatively possibly along the line recently suggested in  \cite{Ji:2020baz}.

\subsection{ $e(x)$ to one-loop}

In Ref.~\cite{Burkardt:2001iy}, the authors have shown in the massive quark model to one-loop that $e(x)$ indeed contains the delta function $\delta(x)$. This is consistent with the above observation that the delta function is nonvanishing in general. In  the massive quark model where $|p\rangle$ is a single quark state,  it is appropriate to employ the scale invariant mass for the `hadron' mass $M$ in Eq.~(\ref{e}),
\beq
M=m(\mu)\left(1+\frac{3\alpha_sC_F}{4\pi} \ln \frac{\mu^2}{m^2}\right),
\eeq
where $C_F=(N_c^2-1)/2N_c$.
The result at one-loop is 
\beq
e(x,\mu)=\delta(1-x) + \frac{\alpha_s}{2\pi} C_F\ln \frac{\mu^2}{m^2}  \left( \frac{2}{[1-x]_+} + \delta(x) + \frac{1}{2}\delta(1-x)\right) ,\label{del}
\eeq  
where $\mu$ is the renormalization scale. As observed in \cite{Burkardt:2001iy}, without the delta function the sum rule 
\beq
\int dx\ e(x) = \frac{\langle p|\bar{\psi}\psi |p\rangle}{2M} = \frac{\partial M}{\partial m},
\eeq
cannot be satisfied. 
Eq.~(\ref{del}) is derived from the following one-loop integral in the light-cone gauge $n\cdot A=A^+=0$ in $d=4-2\epsilon$ dimensions %{\color{red}[$+i\epsilon$ used for both the UV and light-cone singularity regulator. Change notation?]}
\beq
e(x) \sim -2iC_Fg^2 p^+ \int \frac{dk^- d^{d-2}k_\perp}{(2\pi)^d}
% \left[ 
\frac{ (1-\epsilon)(p-k)^2+2\frac{k^2-m^2x}{1-x}   }{(k^2-m^2+i\epsilon)^2((p-k)^2+i\epsilon)} 
%+  \frac{ (1-\epsilon)(k^2+m^2-2p\cdot k)-2\frac{k^2+m^2x}{1+x}    }{(k^2-m^2+i\epsilon)^2((p+k)^2+i\epsilon)}  \right]
, \label{on}
\eeq
where $x=k^+/p^+$. [We use the same letter $\epsilon$ for the small dimension in dimensional regularization  and in the $i\epsilon$ prescription of the propagator, but the distinction should be obvious.] 
The first term in the numerator is proportional to
\beq
\int dk^-  \frac{1}{(k^2-m^2+i\epsilon)^2} = \frac{i\pi\delta(k^+)}{k_\perp^2+m^2},\label{del2}
\eeq 
which is the origin of the delta function $\delta(x)$ in Eq.~(\ref{del}).  
%From the pole structure of the denominator
%\beq
%\frac{1}{(2xP^+k^- -k_\perp^2 + i\epsilon)^2(-2(1-x)P^+k^- + (1-x)m^2-k_\perp^2+i\epsilon)}
%\eeq
%one sees that the second term is nonzero for $1>x>0$ and leads to the $1/[1-x]_+$ term in (\ref{del}). 
%Integrating over $x$, we get 
%\beq
%\left.\int dx e(x)\right|_{conn} = 1+\frac{C_F\alpha}{2\pi}  \left(\frac{1}{\epsilon} -\gamma_E+\ln 4\pi + \ln \frac{\mu^2}{m^2}\right).\nn
%-2ig^2\int \frac{d^4k}{(2\pi)^4} \frac{(1-\epsilon) (k^2+m^2)}{(k^2-m^2+i\epsilon )^2 ((p-k)^2+i\epsilon)} 
%\eeq

Let us consider the same matrix element but now $|p\rangle$ is an on-shell gluon $p^2=0$ with transverse polarization $\varepsilon\cdot p=\varepsilon\cdot n=0$ and $\varepsilon \cdot \varepsilon^*=-\varepsilon_\perp\cdot \varepsilon_\perp^*=-1$. The one-loop diagrams give 
%\beq
%&&\frac{ig^2}{d-2}T_F\int \frac{dk^- d^{d-2}k_\perp}{(2\pi)^d} {\rm Tr}  \left[\frac{(\Slash k +m)\gamma^\mu (\Slash p-\Slash k)\gamma_\mu (\Slash k+m) }{(k^2-m^2+i\epsilon)^2((p-k)^2+i\epsilon)} +\frac{(\Slash k +m)\gamma^\mu (\Slash p+\Slash k)\gamma_\mu (\Slash k+m) }{(k^2-m^2+i\epsilon)^2((p+k)^2+i\epsilon)}  \right]\nn 
%&&= -4ig^2m\int \frac{dk^- d^{d-2}k_\perp}{(2\pi)^d} \left[ \frac{   2k\cdot p -k^2 }{(k^2-m^2+i\epsilon)^2((p-k)^2+i\epsilon)}  +\frac{ 2k\cdot p+k^2 }{(k^2-m^2+i\epsilon)^2((p+k)^2+i\epsilon)}\right] \nn
%&&=0.
%\eeq
\beq
&&-\frac{ip^+g^2T_F}{2m}\int \frac{dk^- d^{d-2}k_\perp}{(2\pi)^d} {\rm Tr}  \left[\frac{(\Slash k +m)\gamma^\mu (\Slash k-\Slash p+m)\gamma^\nu (\Slash k+m) }{(k^2-m^2+i\epsilon)^2((k-p)^2-m^2+i\epsilon)} +\frac{(\Slash k +m)\gamma^\nu (\Slash p+\Slash k+m)\gamma^\mu (\Slash k+m) }{(k^2-m^2+i\epsilon)^2((p+k)^2-m^2+i\epsilon)}  \right]\varepsilon_\mu \varepsilon^*_\nu \nn 
%&&= 4img^2T_F\int \frac{dk^- d^{d-2}k_\perp}{(2\pi)^d} \left[ \frac{ 4\varepsilon\cdot k \varepsilon^*\cdot k-2k\cdot p +k^2 -m^2}{(k^2-m^2+i\epsilon)^2((p-k)^2-m^2+i\epsilon)}  +\frac{ 4\varepsilon\cdot k \varepsilon^*\cdot k +2k\cdot p+k^2-m^2 }{(k^2-m^2+i\epsilon)^2((p+k)^2-m^2+i\epsilon)}\right] \nn
&& =- 2ip^+g^2T_F\int \frac{dk^- d^{d-2}k_\perp}{(2\pi)^d} \left[ \frac{ \frac{2}{1-\epsilon}k_\perp^2 +(p-k)^2-m^2 }{(k^2-m^2+i\epsilon)^2((p-k)^2-m^2+i\epsilon)}  +\frac{ \frac{2}{1-\epsilon} k_\perp^2+(p+k)^2-m^2 }{(k^2-m^2+i\epsilon)^2((p+k)^2-m^2+i\epsilon)}\right] \nn 
&& = \frac{\alpha_s T_F }{\pi } \Gamma(\epsilon)\left(\frac{\mu^2}{m^2}\right)^\epsilon \Bigl( \delta(x)- (1-x) \Theta(1>x>0)  -( 1+x) \Theta(0>x>-1)  \Bigr),
\eeq
where $T_F=1/2$ and $\Theta(1>x>0)$ denotes a step function which has support on $1>x>0$. The delta function $\delta(x)$ arises from the same integral in Eq.~(\ref{del2}). 
Integrating over $x$, we get zero. This is consistent with the fact that the local operator $\bar{\psi}\psi$ does not mix with $F^{\mu\nu}F_{\mu\nu}$, and the delta function $\delta(x)$ is crucial to ensure this property.   We also see that the mixing does occur  at the level of the $x$-distributions.

\section{Gluon condensate distribution}

Let us now come to the main object of interest. 
With the motivation stated in the introduction, we consider the twist-four distribution
\beq
F(x) =\frac{P^+ }{2M^2}\int \frac{dz^-}{2\pi} e^{ixP^+z^-}\langle P|F_{\mu\nu}(0)W[0,z]F^{\mu\nu}(z^-)|P\rangle, \label{f(x)}
\eeq
Related distributions have been recently introduced in \cite{Ji:2020baz}, but their properties have not been investigated. In this and the next sections, we provide the first analysis of Eq.~(\ref{f(x)}) based on the equation of motion and one-loop calculations. 

The first moment of $F(x)$ is the gluon condensate in the proton
\beq
\int dx F(x) = \frac{1}{2M^2}\langle P|F^{\mu\nu}F_{\mu\nu}|P\rangle.
\eeq 
The second moment vanishes $\int dx\ x F(x)=0$ because $F(x)$ is an even function in $x$. 
Similarly to $e(x)$, and as conjectured in \cite{Ji:2020baz}, we expect that $F(x)$ also has a delta function piece
\beq
F(x) = F_{reg}(x) + \delta(x) {\cal C}. \label{c}
\eeq
To obtain insights into the structure of $F(x)$, consider the following operator relation
\beq
\frac{\partial}{\partial z^-} F_{\mu\nu}(0)W[0,z]F^{\mu\nu}(z^-) &=& F_{\mu\nu}(0)W[0,z]D^+F^{\mu\nu}(z^-) %\nn
%&=& - F_{\mu\nu}(0)W[0,z](D^\mu F^{\nu+}(z^-) + D^\nu F^{+\mu}(z^-)  )
 \nn &=& - 2F_{\mu\nu}(0)W[0,z]D^\mu F^{\nu+}(z^-)  \nn 
&=& 2 F_{\mu\nu}\overleftarrow{D}^\mu WF^{\nu+} -2{\cal D}^\mu (F_{\mu\nu}WF^{\nu+})  \nn &&  -2i\int _0^{z^-}d\omega^- 
F_{\mu\nu}(0)WgF^{+\mu}(\omega^-)WF^{\nu+}(z^-)  ,
\eeq
where we used the Bianchi identity and  ${\cal D}^\mu$ represents the translation operator: ${\cal D}_\mu {\cal O}(0,z^-)  \equiv {\rm lim}_{a\to 0} \frac{1}{a^\mu} ({\cal O}(a,z+a)-{\cal O}(0,z))$.
 %We then use the formula
%\beq
% -F_{\mu\nu}(0)W[0,z]D^\mu F^{\nu+}(z^-) &=&  F_{\mu\nu}\overleftarrow{D}^\mu WF^{\nu+} -{\cal D}^\mu (F_{\mu\nu}WF^{\nu+})  \nn &&  -i\int _0^{z^-}d\omega^- 
%F_{\mu\nu}(0)WgF^{+\mu}(\omega^-)WF^{\nu+}(z^-) \nn  &=& g\bar{\psi}(0) W\gamma_\nu F^{\nu+}(z^-)W \psi(0)   -{\cal D}^\mu (F_{\mu\nu}WF^{\nu+})  \nn 
%&&  -i\int _0^{z^-}d\omega^- 
%F_{\mu\nu}(0)WgF^{+\mu}(\omega^-)WF^{\nu+}(z^-),
%\eeq
%where  This leads to the relation
Further using the equation of motion, we immediately obtain 
\beq
 xF(x) &=&  \frac{i}{M^2}\int  \frac{dz^-}{2\pi} e^{ixP^+z^-}\langle P| g\bar{\psi}(0) W\gamma_\nu F^{\nu+}(z^-)W \psi(0) |P\rangle \nn 
&& -\frac{1}{M^2}\int \frac{dz^-}{2\pi} e^{ixP^+z^-} \int _0^{z^-}d\omega^- \langle P|
F_{\mu\nu}(0)WgF^{+\mu}(\omega^-)WF^{+\nu}(z^-) |P\rangle.  \nn
\Longrightarrow  F_{reg}(x) &=&  \frac{i}{xM^2}\int  \frac{dz^-}{2\pi} e^{ixP^+z^-}\langle P| g\bar{\psi}(0) W\gamma_\nu F^{\nu+}(z^-)W \psi(0) |P\rangle \nn 
&& -\frac{1}{xM^2}\int \frac{dz^-}{2\pi} e^{ixP^+z^-} \int _0^{z^-}d\omega^- \langle P|
F_{\mu\nu}(0)WgF^{+\mu}(\omega^-)WF^{+\nu}(z^-) |P\rangle.
\eeq
We shall interpret $\frac{1}{x}$ as the principal value P$\frac{1}{x}$ to be consistent with the property $F(x)=F(-x)$.
Notice that
\beq
\int dx\ x F(x) \propto \langle P| g\bar{\psi}\gamma_\nu F^{\nu+} \psi|P\rangle =0,
\eeq
 because $g\bar{\psi}\gamma_\nu F^{\nu+} \psi =- \partial_\nu T_q^{\nu+}$ is a total derivative. ($T_q^{\mu\nu}$ is the quark part of the energy momentum tensor.) 
%\beq
%\int dx F_{reg}(x) &=& \frac{-1}{2M^2}\int  dz^-\varepsilon(z^-)\langle P| g\bar{q}(0) W\gamma_\nu F^{\nu+}(z^-)W q(0) |P\rangle \nn 
%&& -\frac{i}{2M^2}\int dz^- \varepsilon(z^-) \int _0^{z^-}d\omega^- \langle P|
%F_{\mu\nu}(0)WgF^{+\mu}(\omega^-)WF^{+\nu}(z^-) |P\rangle,   \label{sec}
%\eeq
Thus  the coefficient of the delta function is 
\beq
{\cal C}&=& \frac{1}{2M^2}\langle P|F^{\mu\nu}F_{\mu\nu}|P\rangle + \frac{1}{2M^2}\int  dz^-\varepsilon(z^-)\langle P| g\bar{\psi}(0) W\gamma_\nu F^{\nu+}(z^-)W \psi(0) |P\rangle \nn 
&& +\frac{i}{2M^2}\int dz^- \varepsilon(z^-) \int _0^{z^-}d\omega^- \langle P|
F_{\mu\nu}(0)WgF^{+\mu}(\omega^-)WF^{+\nu}(z^-) |P\rangle.  \label{we}
\eeq

However,  the recent work \cite{Ma:2020kjz}  suggests that ${\cal C}$ may actually be zero, or at least  there is  a significant cancellation among the three terms in ${\cal C}$.  
From the equation of motion
\beq
D^+ F_{+-}+D^iF_{i-}=gJ_-, \qquad D^+F_{+i}+D^jF_{ji}+D^-F_{-i}= gJ_i,
\eeq
one can formally write 
\beq
F_{+-}= \frac{1}{D^+}(gJ_- - D^iF_{i-}) , \qquad F_{+i} = \frac{1}{D^+} (gJ_i - D^jF_{ji} - D^-F_{-i}). \label{nai}
\eeq
Therefore,
\beq
F^{\mu\nu}F_{\mu\nu} 
&=& 2F^{+-}F_{+-} +2F^{+i}F_{+i} +2F^{-i}F_{-i}+ F^{ij}F_{ij} \nn 
%&=&2F^{+-} \frac{1}{D^+}(gJ_- - D^iF_{i-}) +2F^{+i} \frac{1}{D^+} (gJ_i - D^jF_{ji} -D^-F_{-i}) +2F^{-i}F_{-i}+ F^{ij}F_{ij} \nn
&=& 2F^{+\nu} \frac{1}{D^+} gJ_\nu - 2F^{+-} \frac{1}{D^+}D^iF_{i-} -2F^{+i}\frac{1}{D^+} D^j F_{ji} -2F^{+i}\frac{1}{D^+} D^-F_{-i} \nn&& +2F^{-i}F_{-i}+ F^{ij}F_{ij}  . \label{ff}
\eeq 
The first term on the right hand side can be written as, after taking the forward matrix element   $\langle P|...|P\rangle$ and using translational symmetry,
\beq
-F^{+\nu}(0) \int dz^- \varepsilon(z^-) gJ_\nu(z^-) \to -\int dz^- \varepsilon(z^-) F^{\nu+}(z^-) gJ_\nu(0).
\eeq
 This exactly cancels the second term of Eq.~(\ref{we}). In Appendix we show that the remaining terms in Eq.~(\ref{ff}) exactly cancel the third term of Eq.~(\ref{we}). 
Naively, it thus  seems that the coefficient of the delta function in Eq.~(\ref{we}) vanishes identically. However, again this is  inconclusive.  As in Eq.~(\ref{eom}), one can add an `integration constant'  in Eq.~(\ref{nai})
\beq
F_{+-}=\frac{1}{D^+} (gJ_- -D^i F_{i-} )+ \int dz'^- W[z,z']F_{+-}^0(z'),
\eeq
and similarly for $F_{+i}$. The zero modes 
 $\int F_{+-}^0,\int F_{+i}^0$ are not constrained by the equation of motion and should be regarded as  independent degrees of freedom. We thus expect that, in general, the cancellation is incomplete and there exists a delta function $\delta(x)$ in $F(x)$.

\section{One-loop computation of $F(x)$}

In order to gain insight into the $x$-dependence of $F(x)$, in this section we perform one-loop calculations   for  quark and gluon targets.  We shall be particularly interested in whether $F(x)$ contains the delta function $\delta(x)$ or not.

\subsection{Quark target}

We use the light-cone gauge $n\cdot A=A^+=0$ to eliminate the Wilson line. The gluon propagator is proportional to the tensor
\beq
g^{\mu\nu}-\frac{k^\mu n^\nu+k^\nu n^\mu}{k\cdot n}.
\eeq
 We specify the prescription for the pole $1/k\cdot n$ when need arises. 
For an on-shell quark external state $p^2=m^2$, we find 
\beq
&&\int \frac{dk^- d^{d-2}k_\perp}{(2\pi)^d} ig^2\bar{u}(p) \left[\frac{\gamma_\alpha ( \Slash p -\Slash k +m)\gamma_\beta}{(k^2+i\epsilon) ((p-k)^2-m^2+i\epsilon)} +\frac{\gamma_\alpha ( \Slash p +\Slash k +m)\gamma_\beta}{(k^2+i\epsilon) ((p+k)^2-m^2+i\epsilon)}  \right] u(p)  \nn 
&& \qquad  \times  2   \left( g_{\alpha\beta}- \frac{n_\alpha k_\beta + n_{\beta} k_\alpha }{n\cdot k} \right)   \nn
%&& =4 ig^2C_F\int \frac{dk^- d^2k_\perp}{(2\pi)^4} \frac{1}{k^2+i\epsilon} \left[ \frac{ 2m^2 +(d-2)p\cdot k  -\frac{2n\cdot p}{n\cdot k} (2p\cdot k -k^2)}{  -2p\cdot k + k^2+i\epsilon}  + ( k \leftrightarrow -k)\right] \nn
%&&=4 ig^2C_F\int \frac{dk^- d^{d-2}k_\perp}{(2\pi)^d} \frac{1}{k^2+i\epsilon} \left[ \frac{ 2m^2 +(d-2)p\cdot k }{  -2p\cdot k + k^2+i\epsilon} + \frac{2n\cdot p}{n\cdot k}  + ( k \leftrightarrow -k)\right] \nn 
%&& =4 ie^2\int \frac{dk^- d^2k_\perp}{(2\pi)^4} \left[ \frac{ 2m^2k^2 +2(1-\epsilon)p\cdot k k^2 +(p\cdot k)(-2p\cdot k+k^2)}{ (k^2+i\epsilon)^2 (-2p\cdot k + k^2+i\epsilon)}  + ( k \leftrightarrow -k)\right] \nn
&& = 4 ig^2C_F \int \frac{dk^- d^{d-2}k_\perp}{(2\pi)^d}  \Biggl[ \frac{2m^2}{(k^2+i\epsilon)((p-k)^2-m^2+i\epsilon) } +\frac{1-\epsilon}{(p-k)^2-m^2+i\epsilon} -\frac{1-\epsilon}{k^2+i\epsilon} +  ( k \leftrightarrow -k)
\Biggr] ,
 \label{int}
\eeq
where $k^+=xp^+$. % and we used the relation $2p\cdot k = -(p-k)^2+m^2+k^2=(p+k)^2-m^2-k^2$. 
Note that the pole $1/n\cdot k$ has canceled  between the two diagrams.
The first term  on the last line of Eq.~(\ref{int}) is nonvanishing when $1>x>0$ and can be evaluated in a standard manner.
%\beq
%\int \frac{dk^-d^{d-2}k_\perp}{(2\pi)^d} \frac{1}{ (k^2+i\epsilon)((p-k)^2-m^2+i\epsilon)} 
% = \frac{i}{16\pi^2 p^+}\Gamma(\epsilon)\left(\frac{\mu^2}{x^2m^2}\right)^\epsilon. \label{d3} 
%\eeq
%\beq
%\frac{- 8\alpha_s C_Fm^2}{p^+} \int \frac{d^{2-2\epsilon}k_\perp}{(2\pi)^{2-2\epsilon}}  \frac{1}{k_\perp^2+x^2m^2} = -\frac{\alpha_sC_F}{\pi } \frac{2m^2}{p^+} \left(\frac{1}{\epsilon} %-\gamma_E + \ln 4\pi
% + \ln \frac{\mu^2}{x^2m^2}\right).
%\eeq
The second term is proportional to the delta function at $x=1$,
\beq
 4ig^2C_F \int \frac{dk^- d^{d-2}k_\perp}{(2\pi)^d}  \frac{ 1-\epsilon }{-2(1-x)p^+k^- -k_\perp^2-xm^2+i\epsilon}   =  -\frac{\alpha_s C_F m^2}{\pi p^+} \Gamma(\epsilon) \left(\frac{\mu^2}{m^2}\right)^\epsilon  \delta(1-x).
\eeq
%Integrating over $x=k^+/p^+$, we get
%\beq
%C= \frac{4(1-\epsilon)ig^2C_F}{p^+} \int \frac{d^4k}{(2\pi)^4}\frac{1}{(p-k)^2-m^2+i\epsilon} =-\frac{\alpha_s C_F m^2}{\pi p^+} \Gamma(\epsilon) \left(\frac{\mu^2}{m^2}\right)^\epsilon.
%\eeq
%There are two poles at  
%\beq
%k^- = \frac{k_\perp^2}{2xp^+}\equiv a, \qquad k^- = -\frac{k_\perp^2 + xm^2}{2(1-x)p^+}\equiv b
%\eeq
The third term vanishes. We thus find, for $1\ge x \ge -1$, 
\beq
F(x) = -\frac{\alpha_s C_F}{\pi} \left(1+\frac{1}{2}\delta(1-x) + \frac{1}{2}\delta(1+x)\right)\left(\frac{1}{\epsilon} + \ln \frac{\mu^2}{x^2m^2}\right),
\eeq 
and consequently, 
\beq
\int_{-1}^1 dx F(x) =- \frac{3\alpha_s C_F}{\pi}\left(\frac{1}{\epsilon} + \ln \frac{\mu^2}{m^2}  +\frac{4}{3}\right) . \label{ob}
%=  -2\gamma_{m0}\alpha_s \left(\ln \frac{\mu^2}{m^2}+\frac{4}{3}\right).
\eeq
%As a consistency check, let us return to the last line of (\ref{int}) and first integrate over $x=k^+/p^+$. The two terms give an equal contribution  
%\beq
%2\times \frac{4ig^2C_F}{p^+} \int \frac{d^dk}{(2\pi)^d} \frac{2m^2+2p\cdot k(1-\epsilon)}{k^2((p-k)^2-m^2)}
%&=& -\frac{g^2C_F}{2\pi^2 p^+} \int_0^1 dy 2m^2(1+y(1-\epsilon)) \left(\frac{1}{\epsilon}+\ln \frac{\mu^2}{y^2m^2} \right)\nn 
%&=& -\frac{2m^2}{p^+} \frac{3\alpha_s C_F}{\pi}\left(\frac{1}{\epsilon} +\ln \frac{\mu^2}{m^2}+\frac{4}{3}\right) ,
%\eeq 
%in agreement with (\ref{ob}) including the finite part. 

Eq.~(\ref{ob}) is the expected result consistent with the known operator relation
\beq
(F^2)_0 &=&\left(1
+  \beta_0 \frac{\alpha_s}{4\pi\epsilon}   \right)F^2 -\frac{2\gamma_{m0}\alpha_s}{\epsilon} m\bar{\psi}\psi ,
\label{ren} %\nn
%(\bar{\psi}\psi)_0 &=& \left(1+\frac{\gamma_{m0}\alpha_s}{2\epsilon}\right) \bar{\psi}\psi ,
\eeq
where the left hand side is the bare operator. $\beta_0= \frac{11N_c}{3}-\frac{4n_fT_F}{3}$ and $\gamma_{m0}=\frac{3C_F}{2\pi}$ is the first coefficient of the mass anomalous dimension  $\gamma_m =\gamma_{m0}\alpha_s+\cdots$.   Our result gives an interesting new perspective on this well-known result in Eq.~(\ref{ren}). The one-loop anomalous dimension $\gamma_{m0}$ originates from the delta function spikes at $x=\pm 1$ (meaning that the gluon carries away all the quark's energy) and an almost flat distribution for $1>x>-1$. Curiously, the delta function $\delta(x)$ is absent, in contrast to $e(x)$ in the same model.   In the next subsection we perform the same analysis for the coefficient of $F^2$ in Eq.~(\ref{ren}). 

\subsection{Gluon target}

\begin{figure}[t]
	\centering
	
	\begin{subfigure}{0.2\textwidth}
  		\includegraphics[width=\linewidth]{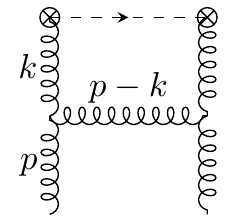}
  		\caption{}
  	\end{subfigure}
  \hspace{2cm}
 	\begin{subfigure}{0.5\textwidth}
 		\includegraphics[width=\linewidth]{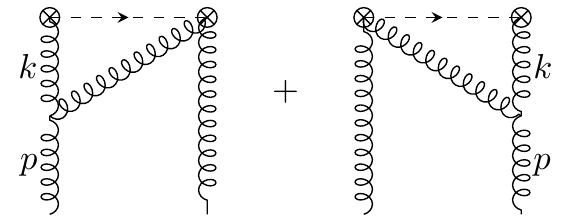}
  		\caption{}
 	\end{subfigure}
 
 	\begin{subfigure}{0.6\textwidth}
 		\includegraphics[width=\linewidth]{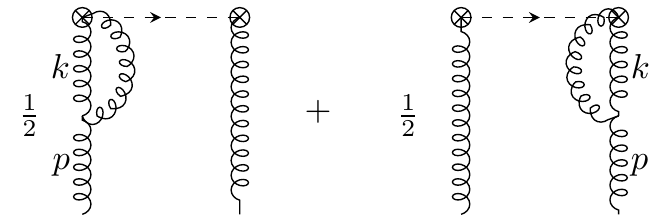}
   		\caption{}
 	\end{subfigure}

  \caption{Feyman diagrams that contribute to $F(x)$ in the light-cone gauge. The dashed line denotes the Wilson line which is set to unity in this gauge. The self-energy diagrams are omitted. }
  \label{diagrams}
\end{figure}

Next we consider the case where the target $|p\rangle$ is a single gluon with transverse polarization. To regularize the infrared divergence, the gluon is assumed to be off-shell with spacelike momentum $p^2=2p^+p^-< 0$. Accordingly, we take $M^2=-p^2$.   % As before, the polarization vector is assumed to be transverse  $\varepsilon(p)\cdot p=\varepsilon(p)\cdot n=0$.   
To zeroth order
\beq
F(x)= \delta(1-x)+\delta(1+x).
\eeq
To one-loop, the diagrams which give nonvanishing contributions are listed in Fig.~\ref{diagrams}.  There are also the self-energy diagrams to be considered later. 
After straightforward calculations we find, for $1>x=k^+/p^+>0$,\\
Fig.~1(a)+(b):
\beq
%&&  ig^2N_c \int \frac{dk^-d^{d-2}k_\perp}{(2\pi)^d} 
%\frac{-8x\varepsilon\cdot k \varepsilon^*\cdot k-2(x^2+5x-4)(p-k)^2-8(1-x)^2p^2 -2x(1-x)k^2}{x(1-x)(k^2+i\epsilon)((p-k)^2+i\epsilon)}  \\ 
%&& =
\frac{p^+}{-2p^2}  ig^2N_c \int \frac{dk^-d^{d-2}k_\perp}{(2\pi)^d} 
\frac{-\frac{4xk_\perp^2}{1-\epsilon} -2(x^2+5x-4)(p-k)^2-8(1-x)^2p^2-2x(1-x)k^2 }{x(1-x)(k^2+i\epsilon)((p-k)^2+i\epsilon)} .  \label{ab} 
\eeq
Fig.~1(c):
%\beq
%&&\frac{1}{2}\times 2\times ip^2 \frac{g^2N_c}{d-2} \delta(1-x)  \int \frac{d^dk}{(2\pi)^d} \frac{-4 (3k^+p^+-3(k^+)^2-2(p^+)^2)
%+2xk^2 
%}{k^+(p^+-k^+)(k^2+i\epsilon)((p-k)^2+i\epsilon)}   \nn
%&&= p^2\frac{\alpha_s N_c}{\pi }\delta(1-x)   \int^1_0 dx' \frac{ \frac{3}{2}x'(1-x')-1 +\frac{1}{2}\epsilon x'(x'-1) }{x'^{1+\epsilon}(1-x')^{1+\epsilon}} \left(\frac{1}{\epsilon}+ \ln \frac{\mu^2}{-p^2}\right)
%\eeq
%\beq
%&&\frac{1}{2}\times 2\times ip^2 g^2N_c \delta(1-x)  \int \frac{dx' dk^- d^{d-2}k_\perp}{(2\pi)^d} \frac{-8x'(1-x')+4
%}{x'(1-x')(k^2+i\epsilon)((p-k)^2+i\epsilon)}   
%\eeq
\beq
&& \frac{p^+}{-2p^2} ig^2N_c \delta(1-x)  \int \frac{dx' dk^- d^{d-2}k_\perp}{(2\pi)^d} \frac{\frac{2k_\perp^2}{1-\epsilon} +2(3x'-2)(p-k)^2 -2(3x'-1)k^2+4(1-2x'(1-x'))p^2
}{x'(1-x')(k^2+i\epsilon)((p-k)^2+i\epsilon)}   . \label{cc}
\eeq
where $x'=k^+/p^+$. The result for $x<0$ is simply obtained by $x \to -x$, $k^\mu \to -k^\mu$. 

At this point we must specify the prescription for the spurious poles $1/k^+\sim 1/x$ and $1/(p^+-k^+)\sim 1/(1-x)$.
If one uses the principal value (pv) prescription 
\beq
\frac{1}{[k^+]_{\rm pv}}=\lim_{\delta\to 0}\frac{k^+}{(k^+)^2+\delta^2 }, \qquad \frac{1}{[p^+-k^+]_{\rm pv}} = \lim_{\delta\to 0}\frac{p^+-k^+}{(p^+-k^+)^2+\delta^2},
\eeq
the $k^-$ integral does not interfere with the poles.  Then the terms proportional to $(p-k)^2$ and $k^2$ in the numerator can be dropped. However, the remaining integrals contain frame-dependent divergences $\sim \ln p^+/\delta$ whose cancellation is nontrivial. This is a well-known symptom of the principal value prescription. Here we  instead  adopt 
 the Mandelstam-Leibbrandt (ML) prescription \cite{Leibbrandt:1987qv},
\beq
\frac{1}{[k^+]_{\rm ML}}=\frac{1}{k^++i\epsilon k^-}, \qquad \frac{1}{[p^+-k^+]_{\rm ML}} = \frac{1}{p^+-k^++i\epsilon(p^--k^-)}.
\eeq
With this choice, one can write 
\beq
\frac{1}{[k^+]_{\rm ML} [p^+-k^+]_{\rm ML}} = \frac{1}{p^+} \left( \frac{1}{[k^+]_{\rm ML}} + \frac{1}{[p^+-k^+]_{\rm ML}} \right),
\eeq
and use the master integrals collected in Appendix B. 
%(\ref{ab})  as
%\beq
% \frac{\alpha_s N_c}{2\pi } \left[ \left(\frac{1}{\epsilon} + \ln \frac{\mu^2}{-p^2}\right) \left(2-x - \frac{2}{[x]_+}    \right)  -x  +\ln [x(1-x)] +4 +\frac{2}{[x]_+} \ln (1-x)+ \delta(1-x) \right]
%\eeq
%and similarly (\ref{cc}) becomes 
%\beq
% \frac{\alpha_s N_c}{2\pi }  \frac{-3\delta(1-x)}{2}  \left(\frac{1}{\epsilon} + \ln \frac{\mu^2}{-p^2} +\frac{7}{3} -\frac{2\pi^2}{9}\right). 
%\eeq
The result for the total contribution from the three diagrams is
\beq
  {\rm (a)+(b)+(c)} &=&\frac{\alpha_s N_c}{2\pi } \Biggl[ \left(\frac{1}{\epsilon} + \ln \frac{\mu^2}{-p^2}\right) \left(2-x - \frac{2}{[x]_+} -\frac{3}{2}\delta(1-x)   \right) \nn 
 && \qquad -x+(x-2)\ln x(1-x) +\frac{2\ln(1-x)}{x} +\left(\frac{\pi^2}{3} -\frac{5}{2}\right)\delta(1-x) \Biggr], \label{sum}
\eeq
%\beq
 % {\rm (a)+(b)+(c)} &=&\frac{\alpha_s N_c}{2\pi } \Biggl[ \left(\frac{1}{\epsilon} + \ln \frac{\mu^2}{-p^2}\right) \left(2-x - \frac{2}{[x]_+} -\delta(1-x)  -\frac{1}{2}\delta(x)  \right) \nn 
% && \qquad -x+(x-2)\ln x(1-x) +\frac{2\ln(1-x)}{x} +\left(\frac{\pi^2}{3} -2\right)\delta(1-x) - \frac{1}{2}\delta(x)\Biggr],\label{sum}
%\eeq
where the plus-prescription $1/[x]_+$ is defined as
\beq
\frac{1}{[x]_+}\equiv  \frac{1}{x^{1+\epsilon}} - \delta(x)\int_0^1 \frac{dx'}{x'^{1+\epsilon}} . \label{plus}
\eeq
 We thus see that, similarly to $e(x)$, $F(x)$ also contains the delta function at $x=0$. However, the way it appears is somewhat unexpected. The coefficient of $\delta(x)$ is divergent, and its only role is to cancel the familiar soft gluon singularity $1/x$ in the first moment. This is a potentially important observation that may find other applications.  
  
 Notice that the $x$-integral of Eq.~(\ref{sum}) vanishes exactly including the finite terms 
$\int_0^1 dx \left( {\rm (a)+(b)+(c)} \right)=0$. This is a special feature of  the ML prescription which is not shared  by the principal value prescription. It actually agrees with the result obtained in the background field gauge  \cite{Tarrach:1981bi} (for the divergent part), namely, the renormalization of the local operator $F^{\mu\nu}F_{\mu\nu}$ solely comes from the self-energy insertion into the external legs. However, in the $x$-space we find an interesting redistribution of partons. The finite part (obtained after removing the $1/\epsilon$ pole and setting $\mu^2=-p^2$) is plotted in Fig.~2. The density of $F^2$  is negative in the large-$x$ region $1>x>0.3$, and this depletion is exactly compensated by the positive region at small-$x$ and the  delta functions at $x=1$.

\begin{figure}
  \includegraphics[width=0.4\linewidth]{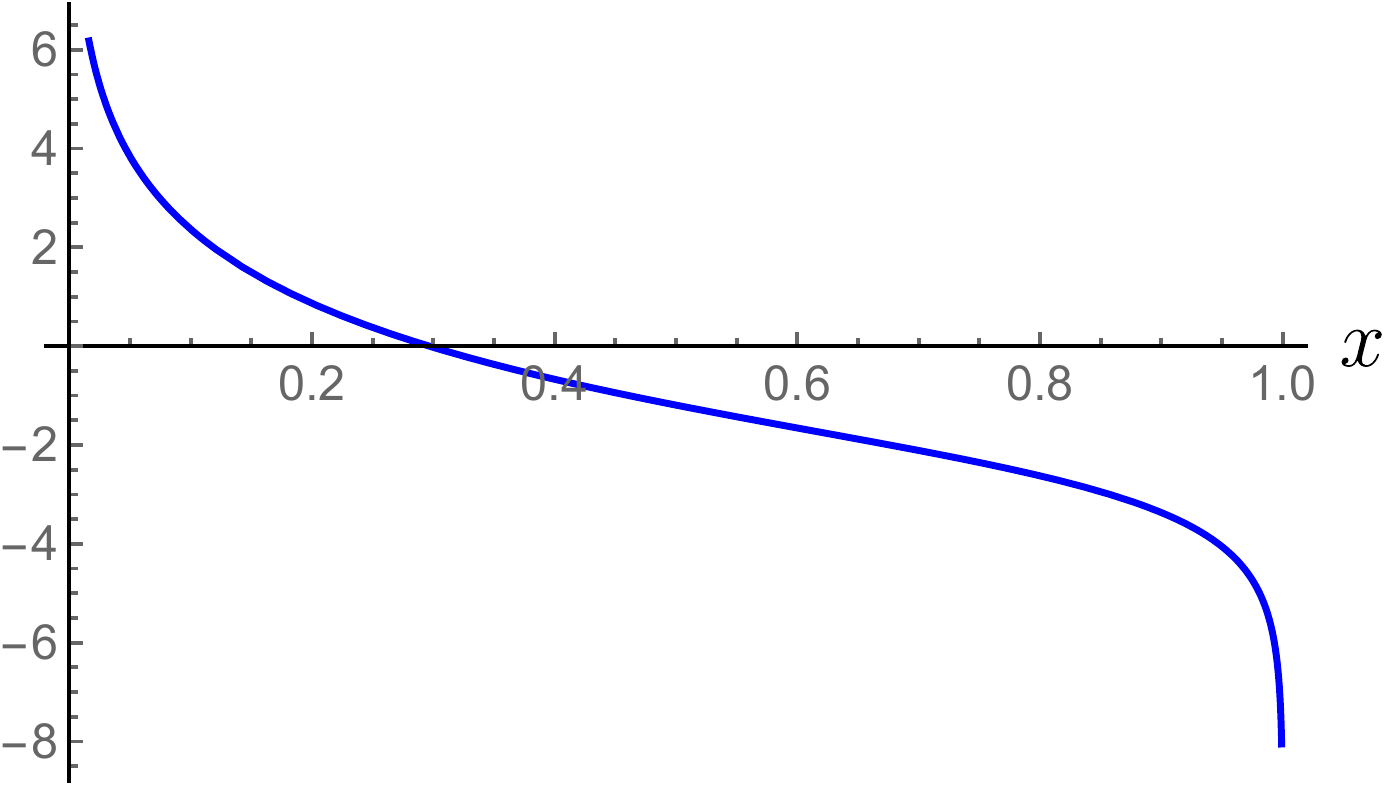}
  \caption{Plot of the function $-x+(x-2)\ln x(1-x) +\frac{2\ln(1-x)}{x}$, see Eq.~(\ref{sum}), for $1>x>0$.  }
  \label{2}
\end{figure}

 The self-energy diagrams modify the leading term as, again in  the ML prescription \cite{Dalbosco:1986eb},
\beq
\delta(1-x) \to \delta(1-x) \left[1+\frac{\alpha_s N_c}{2\pi } \left( \frac{\beta_0}{2N_c\epsilon}-\frac{\pi^2}{3}+\frac{67}{18}  -\frac{5n_f}{9N_c}\right) \right] .
\eeq  
Adding all  contributions, we arrive at, for $1>x>0$, 
\beq
F(x)= \delta(1-x)&+&\frac{\alpha_s N_c}{2\pi} \left[2-x - \frac{2}{[x]_+} + \left(-\frac{3}{2}+ \frac{\beta_0}{2N_c}\right) \delta(1-x)  \right] \left(\frac{1}{\epsilon} + \ln \frac{\mu^2}{-p^2} \right) \nn
&& \qquad + \frac{\alpha_s N_c}{2\pi} \left[     -x+(x-2)\ln x(1-x) +\frac{2\ln(1-x)}{x} +\left(\frac{11}{9} -\frac{5n_f}{9N_c}\right)\delta(1-x)     \right].
\eeq
%\beq
%F(x)= \delta(1-x)&+&\frac{\alpha_s N_c}{2\pi} \left\{2-x - \frac{2}{[x]_+} + \left(-1+ \frac{\beta_0}{2N_c}\right) \delta(1-x)  -\frac{1}{2}\delta(x) \right\} \left(\frac{1}{\epsilon} + \ln \frac{\mu^2}{-p^2} \right) \nn
%&& \qquad + \frac{\alpha_s N_c}{2\pi} \left[     -x+(x-2)\ln x(1-x) +\frac{2\ln(1-x)}{x} +\left(\frac{31}{18} -\frac{5n_f}{9N_c}\right)\delta(1-x)    -\frac{1}{2}\delta(x)   \right].
%\eeq
The result for $0>x>-1$ is simply given by $F(x)=F(-x)$. 
The first moment reads
\beq
\frac{1}{2}\int_{-1}^1 dx F(x) =1+ \beta_0\frac{\alpha_s}{4\pi} \left(\frac{1}{\epsilon} + \ln \frac{\mu^2}{-p^2} \right)  + \frac{\alpha_s N_c}{2\pi } \left(-\frac{\pi^2}{3}+\frac{67}{18}  -\frac{5n_f}{9N_c}\right)  ,
% + \frac{\alpha_s N_c}{2\pi} \left( \frac{49}{18} -\frac{5n_f}{9N_c} -\frac{\pi^2}{3} \right) ,
\eeq
%The second moment is zero because $F(x)$ is an even function. The third moment gives 
%\beq
%\int_{-1}^1 dx F(x) = 2 \left( 1 + \left(\frac{7}{2}-\frac{n_f}{18}\right)\frac{\alpha_s N_c}{\pi } \right)
%\eeq
in agreement with Eq.~(\ref{ren}). 
Incidentally, the $n$-th moment is given by, for even $n\ge 2$, 
\beq
\frac{1}{2}\int_{-1}^1 dx\ x^n F(x) = 1+ \frac{\alpha_s N_c}{2\pi} \left(-\frac{n^2+3n+4}{n(n+1)(n+2)} - \frac{3}{2} + \frac{\beta_0}{2N_c}\right)\frac{1}{\epsilon}+\cdots.
\eeq
It is tempting to relate this result to the anomalous dimension of the operator $F^{\mu\nu}(D^+)^nF_{\mu\nu}$. However, this is nontrivial because for high-dimension operators one has to compute multi-point Greens' function, not just the two-point function, in order to disentangle the mixing with other operators. A proper treatment in the case of $e(x)$ has been given in \cite{Burkardt:2001iy}. Yet, very little is known about the anomalous dimension of high-dimensional, higher-twist gluonic operators \cite{Gracey:2002he,Kim:2015ywa}. 
We leave this to future work.

\section{Conclusions}

In this paper we have introduced the twist-four parton distribution function $F(x)$ which integrates to  the gluon condensate $\langle P|F^2|P\rangle$ and studied its properties based on the equation of motion relations and one-loop calculations. 
Our work literally adds a new dimension---momentum fraction $x$---to the study of nucleon mass structure.  In the future, it would be interesting to further include the dependence on the transverse momentum  $F(x,k_\perp)$ as was done for the quark distribution $e(x,k_\perp)$ (see e.g., Ref.~\cite{Pasquini:2018oyz}).  However, 
at the moment, all this is highly formal and mostly of conceptual interest. The first moment $\langle P|F^2|P\rangle$ can be probed in near-threshold quarkonium production 
\cite{Kharzeev:1998bz,Hatta:2018ina,Boussarie:2020vmu}, but identifying experimental processes that are sensitive to the $x$-dependence will be more challenging. 

Both the operator analysis and one-loop calculations  suggest that  $F(x)$ contains the delta function $\delta(x)$. After all, this is  physically reasonable and could have been anticipated since the zero mode $x=0$ is the genuine nonperturbative sector of light-front quantization \cite{Nakanishi:1976vf,Yamawaki:1998cy,Brodsky:1997de}, and therefore it has to do with the generation of hadron masses. In perturbation theory, there is of course no issue of mass generation. Still,  the delta function is necessary for the consistency of the calculation, like reproducing the correct anomalous dimension as we have seen and restoring Lorentz invariance as emphasized elsewhere (see, e.g., \cite{Aslan:2018tff} for a recent discussion).  Finally, we emphasize that the structure at  finite $1>x>0$ is equally interesting and has a better chance to be explored either  experimentally or in lattice QCD. In particular, we predict the enhancement at small-$x$ due to the familiar soft gluon divergence  in QCD. It would be interesting to study  higher order corrections to this behavior (for example along the line of \cite{Levin:1992mu,Bartels:1999xt})  and also the possible impact of the gluon saturation. 

 % Finally,  the evolution of $F(x)$ is expected to be  complicated and deserves further study. While the first moment correctly reproduces the known anomalous dimension of the operator $F^2$, consideration of higher moments necessarily involves the issue of operator mixing, see a related discussion for $e(x)$ in \cite{Burkardt:2001iy}. Very little is known about the anomalous dimension of high-dimension, higher-twist gluonic operators \cite{Gracey:2002he,Kim:2015ywa}. 

\section*{Acknowledgments}
We are grateful to Kazuhiro Tanaka for discussions. This work is supported by the U.S. Department of Energy, Office of
Science, Office of Nuclear Physics, under contract No. DE- SC0012704,
and in part by Laboratory Directed Research and Development (LDRD)
funds from Brookhaven Science Associates. Y.~Z. is also partially supported by the U.S. Department of Energy, Office of Science,
Office of Nuclear Physics, within the framework of the TMD Topical Collaboration.

\appendix
\section{Evaluation of Eq.~(\ref{ff}), continued }

The last four terms in Eq.~(\ref{ff}) can be written as,  again assuming translational symmetry, 
\beq
&&F^{+-} (0)\int dz^- \varepsilon(z^-)D^iF_{i-}(z^-) +F^{+i}(0)\int dz^- \varepsilon(z^-)D^j F_{ji}(z^-) +F^{+i}(0)\int dz^- \varepsilon(z^-)D^-F_{-i}(z^-) \nn
 && \qquad  +2F^{-i}F_{-i}+ F^{ij}F_{ij} \nn 
&&= -\int dz^- \varepsilon(z^-)D^iF_{i-}(0)F^{+-} (z^-) -\int dz^-\varepsilon(z^-) D^-F_{-i}(0)F^{+i}(z^-) +2F^{-i}F_{-i} \nn 
&& \qquad -\int dz^-\varepsilon(z^-)D^j F_{ji}(0)F^{+i}(z^-)  + F^{ij}F_{ij} . \label{for} 
\eeq
After integration by parts, the first three terms  of Eq.~(\ref{for}) become 
\beq
 -i\int dz^- \varepsilon(z^-) \int_0^{z^-} dw^-\Bigl(F_{i-}(0)gF^{+i}(w^-) F^{+-}(z^-) +  F_{-i}(0)gF^{+-}(w^-) F^{+i}(z^-) \Bigr) \nn 
+ \int dz^- \varepsilon(z^-)  F_{i-} (D^iF^{+-} -D^-F^{+i})  +2F^{-i}F_{-i} .\label{f1}
\eeq
The second line of Eq.~(\ref{f1}) vanishes because $D^iF^{+-} -D^-F^{+i}=D^+F^{i-}$ so that
\beq
 \int dz^- \varepsilon(z^-)  F_{i-} (0)D^+ F^{i-} (z^-) +2F^{-i}F_{-i} = 0,
\eeq
where we used $\partial_{z^-}\varepsilon(z^-)=2\delta(z^-)$.  
The last two terms of Eq.~(\ref{for}) can be written as 
\beq
-i \int dz^- \varepsilon(z^-) \int_0^{z^-} dw^- F_{ji}(0)gF^{+j}(w^-) F^{+i}(z^-)  +  \int dz^- \varepsilon(z^-) F_{ji}(0)D^j F^{+i}(z^-)  +F^{ij}F_{ij} .\label{th}
\eeq
The last two terms of Eq.~(\ref{th}) actually cancel. This can be seen by writing  $D^jF^{+i}\to \frac{1}{2}(D^jF^{+i}-D^iF^{+i}) = \frac{1}{2}D^+F^{ji}$ and integrating by parts in $z^-$. 
The sum of Eqs.~(\ref{f1}) and (\ref{th}) is then
\beq
 -i\int dz^- \varepsilon(z^-) \int_0^{z^-} dw^-  F_{\mu\nu}(0)gF^{+\mu}(w^-) F^{+\nu}(z^-) .
\eeq
This exactly cancels the second line of Eq.~(\ref{we}).

\section{Useful integrals}

Here we list the integrals needed to evaluate Eqs.~(\ref{ab}) and (\ref{cc}).
\beq
&& \int \frac{d^dk}{(2\pi)^d} \frac{1}{(k^2+i\epsilon)((p-k)^2+i\epsilon)} = \frac{i}{16\pi^2} \Gamma(\epsilon)\left(\frac{\mu^2}{-p^2}\right)^\epsilon   (1+2\epsilon), \\
&&\int \frac{dk^-d^{d-2}k_\perp}{(2\pi)^d} \frac{1}{ (k^2+i\epsilon)((p-k)^2+i\epsilon)} 
%=\frac{i}{2p^+}\int \frac{d^{d-2}k_\perp}{(2\pi)^{d-2}}  \frac{1}{k_\perp^2-x(1-x)p^2}\nn  
= \frac{i}{16\pi^2 p^+}\Gamma(\epsilon)\left( \frac{\mu^2}{-x(1-x)p^2}\right)^{\epsilon} , \label{d1} \\
% &&= \frac{i}{16\pi^2 p^+}\Gamma(\epsilon)\left(\frac{\mu^2}{x^2m^2}\right)^\epsilon.  \nn
&&\int dk^- d^{d-2}k_\perp \frac{1}{[k^+]_{\rm ML} (k^2+i\epsilon)} = 0, \\
&& \int \frac{dk^- d^{d-2}k_\perp}{(2\pi)^d} \frac{1}{[p^+-k^+]_{\rm ML} (k^2+i\epsilon)} = \frac{ip^2}{16\pi^2(p^+)^2}\Gamma(\epsilon)  \left(\frac{\mu^2}{-p^2}\right)^\epsilon \frac{\delta(1-x)}{1-\epsilon}, \label{4}\\
%&& \int \frac{dk^- d^{d-2}k_\perp}{(2\pi)^d} \frac{1}{[k^+]_{\rm ML} ((p-k)^2+i\epsilon)} = \frac{ip^2}{16\pi^2(p^+)^2}\delta(x), \nn
&& \int \frac{dk^- d^{d-2}k_\perp}{(2\pi)^d} \frac{1}{[k^+]_{\rm ML} (k^2+i\epsilon) ((p-k)^2+i\epsilon)} = \frac{i}{16\pi^2(p^+)^2}\Gamma(\epsilon)\left(\frac{\mu^2}{-p^2}\right)^\epsilon \frac{(1-x)^{-\epsilon}}{[x]_+} ,  \label{non}
\\
&& \int \frac{d^d k}{(2\pi)^d} \frac{1}{[k^+]_{\rm ML}(k^2+i\epsilon)((p-k)^2+i\epsilon)} = \frac{i}{96p^+}. \label{last} \\
&& \int \frac{dk^-d^{d-2}k_\perp}{(2\pi)^d} \frac{k_\perp^2}{[p^+-k^+]_{\rm ML} (k^2+i\epsilon) ((p-k)^2+i\epsilon)} = \frac{ip^2}{16\pi (p^+)^2}\Gamma(\epsilon) \left(\frac{\mu^2}{-p^2}\right)^\epsilon \left(\frac{x^{1-\epsilon}}{ (1-x)^\epsilon} -\frac{\delta(1-x)}{1-\epsilon}\right) , \\
&& \int \frac{d^dk}{(2\pi)^d} \frac{k_\perp^2}{[p^+-k^+]_{\rm ML} (k^2+i\epsilon) ((p-k)^2+i\epsilon)} = \frac{-ip^2}{32\pi^2p^+}\Gamma(\epsilon)\left(\frac{\mu^2}{-p^2}\right)^\epsilon (1+{\cal O}(\epsilon^2)).
\eeq
Note that Eq.~(\ref{last}) is finite. The plus-prescription in Eq.~(\ref{non}) is defined in Eq.~(\ref{plus}).  
This can be understood as follows. For $k^+\neq 0$, the prescription is irrelevant and one can use Eq.~(\ref{d1}) to evaluate the integral. On the other hand, the $k^+$-integral of Eq.~(\ref{non}) does not contain divergence due to Eq.~(\ref{last}) so there must be a delta function singularity at $k^+=0$.

\end{document}